\documentclass[apj,a4paper,12pt,useAMS]{emulateapj}

\usepackage{graphicx,graphics}
\usepackage{natbib}

\shorttitle{Magnetars and GRB X-ray afterglows}

\shortauthors{Yu et al.}

\newcommand{\mes}{M\'esz\'aros}

\begin{document}
\title{The role of newly born magnetars in gamma-ray burst X-ray afterglow emission: Energy injection and internal emission}

\author{Yun-Wei Yu\altaffilmark{1,2}}
\author{K. S. Cheng\altaffilmark{1}}
\author{Xiao-Feng Cao\altaffilmark{2}}

\altaffiltext{1}{Department of Physics, The University of Hong Kong,
Pokfulam Road, Hong Kong, China; yuyw@hku.hk, hrspksc@hkucc.hku.hk}
\altaffiltext{2}{Institute of Astrophysics, Huazhong Normal
University, Wuhan, 430079, China}

\begin{abstract}
{\it Swift} observations suggest that the central compact objects of
some gamma-ray bursts (GRBs) could be newly born millisecond
magnetars. Therefore, by considering the spin evolution of the
magnetars against {\it r}-mode instability, we investigate the role
of the magnetars in GRB X-ray afterglow emission. Besides modifying
the conventional energy injection model, we pay particular attention
to the internal X-ray afterglow emission, whose luminosity is
assumed to track the magnetic dipole luminosity of the magentars
with a certain fraction. Following a comparison between the model
and some selected observational samples, we suggest that some
so-called ``canonical" X-ray afterglows including the shallow decay,
normal decay, and steeper-than-normal decay phases could be
internally produced by the magnetars (possibly through some internal
dissipations of the magnetar winds), while the (energized) external
shocks are associated with another type of X-ray afterglows. If this
is true, from those internal X-ray afterglows, we can further
determine the magnetic field strengths and the initial spin periods
of the corresponding magnetars.
\end{abstract}


\slugcomment{2010, ApJ, ??, ??}

\section{Introduction}
Gamma-ray bursts (GRBs) are short, intense flashes of soft
gamma-rays ($\sim0.01-1$ MeV), which are always followed by
long-lasting low-frequency afterglow emission. Usually, the
afterglow emission is attributed to an external forward shock
arising from the interaction of the GRB outflow with the circumburst
medium, whereas the mechanisms responsible for the bursts are under
more debate. Since the launch of the {\it Swift} spacecraft (Gehrels
2004), many detailed features of the GRB X-ray afterglows have been
revealed by the X-Ray Telescope (XRT) aboard. Then Nousek et al.
(2006) and Zhang et al. (2006) phenomenologically summarized a
``canonical" X-ray light curve (LC) with four smooth segments
(sometimes superposed by some sharp flares), although the LCs owning
all components are actually in the tiny minority. Specifically, the
four different emission phases defined by them include: (1) {\it
Initial steep decay phase} that is widely accepted to be the tail of
the prompt emission (i.e., curvature effect; Fenimore et al. 1996;
Kumar \& Panaitescu 2000). (2) {\it Shallow decay (even a plateau)
phase} that is usually ascribed to a continuous energy injection
into the external shock (Rees \& {\mes} 1998; Dai \& Lu 1998a,b;
Zhang \& {\mes} 2001). (3) {\it Normal decay phase} that does not
contradict with the standard external shock model. (4) {\it
Steeper-than-normal decay phase} that is often connected to the jet
break (Rhoads 1997). To summarize, in the conventional picture as
described above, the emission during the shallow decay, normal
decay, and steeper-than-normal decay phases (of interest in this
paper) are usually considered to be associated with the external
shock, while the initial steep decay emission as well as the flares
are probably of internal origin.

The X-ray shallow-decay and flare emission strongly suggest that GRB
central objects should have long activities after the bursts.
Therefore, highly magnetized, rapidly spinning pulsars (i.e.,
millisecond magnetars) gradually become a popular candidate of the
central compact objects (e.g., De Pasquale et al. 2007; Metzger et
al. 2007; Troja et al. 2007; Zhang \& Dai 2008, 2009; Bucciantini et
al. 2009; Corsi \& {\mes} 2009; Lyons et al. 2009), although the
black hole model is still an attractive choice. It is a difficult
task to observationally distinguish between the magnetar and black
hole models. However, some magnetohydrodynamic simulations showed
that the magnetar model could be more mature in the sense that it
provides quantitative explanations for the durations, energies,
Lorentz factors, and collimation of long GRB outflows (Metzger
2010). Moreover, with a spinning-down magnetar, some {\it Swift}-XRT
features can be explained well (e.g., Dai et al. 2006; Fan \& Xu
2006; Yu \& Dai 2007) . Especially, the unusual X-ray afterglow LC
of GRB 070110, where a nearly constant X-ray emission is followed by
a very steep decline of $\alpha\sim 9$ ($\alpha$ is the decline
index of $t^{-\alpha}$), can be understood by ascribing the plateau
emission to magnetar-driven internal emission (Troja et al. 2007).

The internal plateau emission of GRB 070110 also tells us that, in
addition to the initial steep decay and flare phases, the X-ray
emission during any other afterglow phases could also arise from
some internal dissipation mechanisms. Therefore, it is fair to
consider that the long active central objects
of some other GRBs could also play an essential, relatively more
direct role in their afterglow emission, at least in the X-ray band.
In some extreme situations, we suspect that such an internal-origin
emission component could even dominate the total X-ray afterglow
emission of a GRB, whereas the external shock emission is outshined
in the X-ray band. In other words, the conventional external shock
model is only one choice among various afterglow origin models, as
also proposed by some authors before (Ghisellini et al. 2007; Kumar
et al. 2008; Cannizzo \& Gehrels 2009; Lindner et al. 2009; Lyutikov
2009). A wide investigation on the GRB X-ray afterglows given by
Willingale el al. (2007) indeed showed that more than one hundred
X-ray afterglows can be divided into two emission components, one of
which is probably of internal origin.

Following the observational results and the above two theoretical
considerations, in this paper, we investigate in more detail the
role of millisecond magnetars in the X-ray afterglow emission of
some GRBs, based on a careful analysis on the spin evolution of the
magnetars. Besides modifying the conventional energy injection model
(Dai \& Lu 1998a, b; Zhang \& {\mes} 2001), we pay particular
attention to the internal X-ray afterglow emission that is produced
by the magnetars, possibly through some internal dissipations in the
magnetar winds. The luminosity of this internal emission is assumed
to simply track the magnetic dipole luminosity of the magnetars with
a certain fraction. To summarize, the observed X-ray afterglows
could be emitted from two different regions (i.e., an
internally-dissipated magnetar wind and an energized external shock)
at very different radii. The competition between these two emission
components leads to a diversity of the X-ray afterglow LCs.

In Section 2, we briefly review the spin-down of magnetars against
{\it r}-mode instability. We analyze the temporal behaviors of X-ray
afterglows by combining the contributions from a magnetar wind and
an external shock in Section 3, where the energy injection from the
wind to the shock is also taken into account. In Section 4, some
observational samples are selected and fitted in order to confront
the model with observations. Meanwhile, some implications to the
magnetars from the afterglow data are discussed. Finally, a summary
and discussion are given in Section 5.

\section{Spin-down of magnetars}
For long GRBs associated with Type Ibc supernovae, the central
magnetars could be formed during the core-collapse of massive,
rotating stars, since powerful magneto-centrifugal outflows from the
nascent magnetars may stave off black hole formation entirely
(Metzger 2010). For short GRBs, the merger of compact binaries may
also give rise to a massive neutron star if the equation of state of
stellar matter is stiff enough, as implied by the observations of
kilohertz quasi-periodic oscillations in accreting neutron stars
(Klu\'zniak 1997; Klu\'zniak \& Ruderman 1998). So it seems somewhat
acceptable that the central compact objects of some GRBs are
millisecond magnetars. The idea that GRBs could originate from
magnetars actually had been proposed almost two decades ago (Usov
1992, 1994; Paczynski 1992; Duncan \& Thompson 1992), far before the
{\it Swift} era. Specifically, several central engine mechanisms
have been proposed, e.g., a neutrino-driven wind (Thompson 1994;
Metzger et al. 2007; Bucciantini et al. 2009;), a magnetic
reconnection-accelerated wind (Drenkhahn \& Spruit 2002), or an
hyperaccretion onto the magnetars (Zhang \& Dai 2008, 2009). Anyway,
these violent processes lead the initial spin evolution of the
magnetars during the first tens of seconds to be very complicated.

However, on the relatively longer (afterglow) timescales concerned
here, the short-term processes as mentioned above could be no longer
important. Then the spin-down of the magnetars would be mainly due
to electromagnetic torque and the torque connected with
gravitational wave radiation. For the latter, first, a
magnetic-field-caused equatorial ellipticity can generate
gravitational-quadrupole radiation, which however can not provide an
essential contribution to the spin-down unless with an extremely
high magnetic field (Usov 1992). Secondly, a relatively stronger
gravitational wave radiation can be produced through some
nonaxisymmetric stellar perturbations. For nascent neutron stars,
such perturbations can be easily created by {\it r}-mode
instability, which arises from the action of the Coriolis force with
positive feedback (Andersson 1998; Friedman \& Morsink 1998)
succumbing to gravitational radiation-driven
Chandrasekhar-Friedman-Schutz instability.

Following a phenomenological second-order model for the {\it r}-mode
evolution (Owen et al. 1998; S\'a 2004), we can calculate the spin
evolution of a magnetar by (S\'a 2004; Yu et al. 2009a)
\begin{eqnarray}
{dP\over dt} = {4\alpha^2\over15} (\delta+ 2){ P\over \tau_g} + {P
\over \tau_m},
\end{eqnarray}
where $P$ is the spin period of the magnetar, $\alpha$ is the
dimensionless amplitude of the {\it r}-modes, and $\delta$ is a free
parameter describing the initial degree of the differential rotation
of the star. The gravitational and magnetic braking timescales can
be written as $\tau_g=144P_{-3}^6$ s and $\tau_m =
4\times10^5B_{14}^{-2}P_{-3}^2$ s, respectively\footnote{Hereafter,
some basic structural parameters such as the mass, the radius, and
the moment of inertial of the magnetar are taken to be
$1.4M_{\odot}$, $10^6$ cm, and $\rm 10^{45} g~ cm^2$, respectively.
Additionally, the convention $Q_x = Q/10^x$ is adopted in cgs units
and a subscript ``$i$" represents the initial values of the
quantities.}. The viscous damping of the {\it r}-modes is ignored
here due to the high temperatures of $\sim10^{10}$ K in nascent
neutron stars. Correspondingly, the evolution of the {\it r}-mode
amplitude can be calculated from (S\'a 2004; Yu et al. 2009a)
\begin{eqnarray}
{d\alpha\over dt} = \left[1+{2\alpha^2\over15} (\delta+ 2)\right]{
\alpha\over \tau_g} + {\alpha \over 2 \tau_m}.
\end{eqnarray}
In the case of $\tau_{g,i}\gg\tau_{m,i}$ (i.e., $B\gg
B_c=5\times10^{15} P_{i,-3}^{-2}$ G), the spin-down would be
dominated exclusively by the magnetic dipole radiation and the {\it
r}-modes cannot arise sufficiently rapidly. In this case, the spin
evolution can be easily expressed as usual as $P(t) =
P_i(1+t/T_m)^{1/2}$ with $T_m = 2\times10^5B_{14}^{-2} P_{i,-3}^2$
s.

In contrast, for not very high magnetic fields (i.e.,
$\tau_{g,i}\ll\tau_{m,i}$), the spin-down should be first dominated
by the gravitational wave radiation. By ignoring the magnetic term,
Equations. (1) and (2) can be solved analytically and an asymptotic
solution can be written as (S\'a \& Tom\'e 2005, 2006)
\begin{equation}
P(t)\approx\left\{
\begin{array}{ll}
P_i\left[1-{2\over15}\alpha_i^2(\delta+2)\exp(2t/\tau_{g,i})\right]^{-1},&{~\rm for~}t<T_g,\\
1.6P_i(t/\tau_{g,i})^{1/5},&{~\rm for~}t>T_g.
\end{array}\right.\label{pt}
\end{equation}
Here the break time can be solved from $(d^2\alpha/dt^2)_{t=Tg} = 0$
to be $T_g = a\times10^3P_{i,-3}^6$ s, where the prefactor $0.7 < a
< 3.2$ for a wide parameter region of $10^{-10} < \alpha < 10^{-6}$
and $0 < \delta< 10^8$ (Yu et al. 2009b). As the increasing of the
spin period, the magnetic braking effect would eventually exceed the
gravitational braking effect and then the time-dependence of the
spin period changes from $P\propto t^{1/5}$ to $P \propto t^{1/2}$.
We denote the change time by $T_c$, which will be given explicitly
later.

\section{GRB X-ray afterglow emission}
After the bursts, the central remanent magnetars could still keep
abundant rotational energy, although a considerable fraction has
been expanded on the bursts. This remaining energy could be released
peacefully and persistently, and drive a continuous magnetar winds.
We believe that the magnetar winds are probably able to
produce long-lasting emission (i.e., afterglow emission), through
some internal dissipation mechanisms such as magnetic reconnection
(e.g., Giannios \& Spruit 2005) and the terminative shock of the
winds (e.g., Dai 2004; Yu \& Dai 2007) etc. Such an internal-origin
emission component would of course compete with the external shock
emission component, while the latter alone is usually challenged by
some multi-wavelength afterglow observations (e.g., chromatic LC
breaks).

In order to compare with X-ray observations, we first simply assume
that the (isotropically-equivalent) luminosity of the wind X-ray
afterglow emission tracks the magnetic dipole luminosity of the
magnetars with a constant fraction ($\xi$; X-ray radiation
efficiency) as
\begin{eqnarray}
\mathcal L_{X}^{\rm mw}=\xi L_{\rm
md}/f_B=10^{47}\xi_{-1}f_{B,-1}^{-1}L_{\rm md,47}~\rm
erg~s^{-1},\label{lxmw}
\end{eqnarray}
where $f_B = (1-\cos\theta_{w})$ is the beaming factor of the
magnetar winds with $\theta_{w}$ being the half-opening angle of the
winds.
To be specific, the magnetic dipole luminosity of a magnetar can be
calculated by
\begin{eqnarray}
L_{\rm md}(t)={I\Omega^2\over
\tau_m}=10^{47}F(t)B_{14}^2P_{i,-3}^{-4}\rm ~erg~s^{-1},\label{lmd}
\end{eqnarray}
where $I$ is the moment of inertial of the star and $\Omega= 2\pi/P$
is the spin frequency. According to the analysis on the spin-down of
the magnetars, the time-dependence $F(t)$ of the magnetic dipole
luminosity can be expressed approximately as

For $B > B_c$,
\begin{equation}
F(t)\approx\left\{
\begin{array}{ll}
t^0, &t < T_m,\\
\left({t\over T_m}\right)^{-2},&t>T_m;
\end{array}\right.\label{ft1}
\end{equation}

For $B < B_c$,
\begin{equation}
F(t)\approx\left\{
\begin{array}{ll}
t^0, &t < T_g,\\
\left({t\over T_g}\right)^{-q},&T_g<t<T_c,\\
\left({T_c\over T_g}\right)^{-q}\left({t\over
T_c}\right)^{-2},&t>T_c,
\end{array}\right.\label{ft2}
\end{equation}
where $T_c = (T_m^2/T_g^q )^{1/(2-q)}$. Following Equation
(\ref{pt}), the value of $q$ can be taken as 0.8 approximately (a
more exact numerical calculation would show $q\sim 1$). As shown by
Equation (\ref{ft2}) for $B < B_c$, we surprisingly find that the
double-broken power-law behavior of $\mathcal L_{X}^{\rm mw}$ is in
good agreement with the so-called canonical X-ray LC including all
of the shallow decay, normal decay, and steeper-than-normal decay
phases. This may be just a coincidence, but alternatively it seems
also acceptable to ascribe some observed double-broken power-law
X-ray afterglow emission to the magnetar winds rather than the
external shocks. In this case, the change in the slopes of the LCs
is just because of the evolution of the magnetic dipole luminosity,
but independent of the external shock physics. As an analogical
consideration for the magnetars with $B > B_c$, Equation (\ref{ft1})
also predicts some X-ray LCs behaving as a plateau followed by a
steep decay with $\alpha\sim2$.
However, the rare observation of such a type of X-ray LCs implies
that the magnetic field strengths of the GRB magnetars are generally
lower than the critical value $B_c$.\footnote{We will test this
argument in Section 4 and give some related discussion in Section
5.} So we would mainly concern the case of $B < B_c$ in the
following calculations.

Although the internal X-ray afterglow emission is suggested above,
we still think that the external shock could play an important role
in the afterglow emission. In the calculation of shock dynamics, we
also take into account the energy injection into the shock from the
wind as considered before (e.g., Dai \& Lu 1998a, b; Zhang \& {\mes}
2001), even though a part of the wind energy could have been
radiated directly. Following Equations (\ref{lmd}) and (\ref{ft2})
for $B < B_c$, the increasing (isotropically-equivalent) shock
energy can be approximately written as (for $q\leq1$)
\begin{equation}
\mathcal E(t)\approx\mathcal E_{i}\times\left\{
\begin{array}{ll}
t^0, &t < T_{\rm ei},\\
\left({t\over T_{\rm ei}}\right),&T_{\rm ei}<t<T_g,\\
\left({T_g\over T_{\rm ei}}\right)\left({t\over T_g}\right)^{1-q},&T_g<t<T_c,\\
\left({T_g\over T_{\rm ei}}\right)\left({T_c\over
T_g}\right)^{1-q}\equiv \mathcal E_f,&t>T_c.
\end{array}\right.\label{energy}
\end{equation}
The starting time of the energy increase can be calculated by
$T_{\rm ei} = \mathcal E_i/(\zeta f_B^{-1} L_{\rm md,i}) = 100
\zeta_0^{-1}\mathcal E_{i,50}f_{B,-1}L_{\rm md,i,47}^{-1}$ s, where
$\zeta\leq (1-\xi)$ is the fraction of the wind energy that is
injected into the shock. Equation (\ref{energy}) shows that the
energy injection process can be divided into two stages for $q < 1$.
However, since the value of $q$ is actually very close to one, the
energy increase of the shock after $T_g$ is difficult to be detected
and thus can be neglected. Therefore, in comparison with the
conventional energy injection model proposed by Dai \& Lu (1998a, b)
and Zhang \& {\mes} (2001), the main difference here is that the
duration of the energy injection is determined by $T_g$ rather than
$T_m$. For shock dynamic evolution\footnote{Here some shock-related
quantities are defined as follows: $\Gamma$ and $R = 2\Gamma^2ct$
are the Lorentz factor and the radius of the shock, $\mathcal M_{\rm
sw}={4\over 3}\pi R^3nm_p$ is the (isotropically-equivalent) mass of
the medium swept up by the shock, and $n$ is the number density of
the circum medium. $m_p$ and $c$ are the mass of proton and the
speed of light.}, we first introduce the deceleration time of the
shock as $T_{\rm dec} = 3 \mathcal
E^{1/3}_{i,50}\Gamma^{-8/3}_{i,2.3} n^{-1/3}_0$ s ($< T_{\rm ei} <
T_g$).
Before $T_{\rm dec}$, the shock deceleration can be neglected (i.e.,
$\Gamma\sim\Gamma_i$) since $\mathcal M_{\rm sw}$ is insufficiently
high. After $T_{\rm dec}$, the dynamic evolution of the shock can be
obtained from the energy conservation law $\mathcal E =
\Gamma^2\mathcal M_{\rm sw}c^2$ as
\begin{equation}
\Gamma(t)=\left({3\mathcal E\over 32\pi n m_pc^5t^3}\right)^{1/8}
\propto\left\{
\begin{array}{ll}
t^{-3/8},&T_{\rm dec}<t<T_{\rm ei},\\
t^{-1/4},&T_{\rm ei}<t<T_g,\\
t^{-3/8},&t>T_g.
\end{array}\right.
\end{equation}
Obviously, the shock deceleration during $T_{\rm ei} < t < T_{g}$ is
effectively decreased by the energy injection.

For the afterglow emission of interest, here we only concern the
shock emission after $T_{\rm ei}$. Using the above dynamic results
and following Sari et al. (1998), the (isotropically-equivalent)
luminosity of the shock synchrotron X-ray afterglows can be given
analytically by
\begin{equation}
\mathcal L_X^{\rm sh}=\mathcal L_{X,\rm ei}^{\rm sh}\times\left\{
\begin{array}{ll}
\left({t\over T_{\rm ei}}\right)^{-(p-2)/2},&T_{\rm ei}<t<T_g,\\
\left({T_g\over T_{\rm ei}}\right)^{-(p-2)/2}\left({t\over
T_{g}}\right)^{-(3p-2)/4},&t>T_g,\label{lxsh}
\end{array}\right.
\end{equation}
where $p$ is the spectral index of the energy distribution of the
shock-accelerated electrons. The luminosity at $T_{\rm ei}$ reads 
\begin{eqnarray}
\mathcal L_{X,\rm ei}^{\rm sh}
&\approx&10^{46}\left(g_{p,-0.5}\epsilon_{e,-1}\right)^{p-1}\left(\zeta_0
f_{B,-1}^{-1}L_{\rm
md,i,47}\right)^{(3p-2)/4}\nonumber\\
&&\times\left(\nu_{X,17.5}\mathcal
E_{i,50}\right)^{(2-p)/2}\epsilon_{B,-2}^{(p-2)/4}\rm
erg~s^{-1},\label{lxsh0}
\end{eqnarray}
where $g_p = (p -2)/(p-1)$, $\nu_X$ is the X-ray frequency, and
$\epsilon_e$ and $\epsilon_B$ are the equipartition factors of the
electron internal energy and magnetic energy, respectively.

Equation (\ref{lxsh0}) shows that, for $p$ being not much higher
than two, the shock luminosity is mainly sensitive to the parameters
$\epsilon_e$, $\zeta$, $L_{\rm md}$, and $f_{B}$. So an upper limit
for the shock luminosity for $t>T_{\rm dec}$ can be given as follows
\begin{eqnarray}
\mathcal L_{X,\rm upper}^{\rm sh}=\epsilon_e \zeta L_{\rm md,
i}/f_{B}=10^{47}\epsilon_{e,-1}\zeta_{0}f_{B,-1}^{-1}L_{\rm
md,i,47}~\rm erg~s^{-1}.\label{lxshupper}
\end{eqnarray}
A comparison between Equations (\ref{lxmw}) and (\ref{lxshupper})
shows that the shock emission can exceed the wind emission only for
$\epsilon_e\gg\xi/(1-\xi)$. By varying the value of $\xi$ and fixing
$\epsilon_e=0.2$, we plot the wind- and shock-contributed X-ray LCs
in Figure 1. Combining the two emission components, the model in
principle predicts three types of LCs as exhibited, i.e., an
energized shock-dominated type, a magnetar wind-dominated type, and
an intermediate type. In particular, the intermediate type LCs could
have an interesting but complicated profile. It will be a valuable
attempt to find some observational cases with such a profile,
although it would be not easy. For a general investigation, we would
only concern the other two types in this paper. Anyway, it seems
soemwhat too simple that nearly all of the {\it Swift}-XRT
afterglows can be summarized by a canonical LC. This view obscures
the possible diversity of the physical origins of the X-ray
afterglows. In our opinion, we had better treat all of the observed
afterglows as a collective of several different classes,
which corresponds to different afterglow mechanisms.
\section{Confronting the model with observations}
\subsection{Fittings to some observational LCs}
As shown by Figure 1, both the energized shock- and magnetar
wind-dominated afterglow LCs initially have a very flat segment, the
decay index of which is close to be zero if $p$ is not much higher
than two. Therefore, we select observational samples from the public
XRT-team LC repository (Evans et al. 2007, 2009) between 2005
January and 2009 December under four criterions: (i) the data are
rich enough to clearly exhibit the profile of the LC and no flare
appears after the initial steep decay phase; (ii) a remarkable
plateau emission ($\alpha<0.3$) immediately follows the initial
steep decay phase; (iii) the decay index of the last segment of the
LC is not much larger than 2; and (iv) the GRB's redshift is known.
As a result, 16 representative samples are obtained, which naturally
fall into two classes as shown in Figures 2 and 3. The former LCs
have only one break after the initial steep decay phase, while the
latter LCs have two breaks. Using the following smoothed broken and
double-broken power law functions ($w = 3$; Liang et al. 2007),
\begin{eqnarray}
F_X(t) &=& F_{X,b}^{(1)}\left[\left({t\over
T^{(1)}_{b}}\right)^{w\alpha^{(1)}_1} + \left({t\over
T^{(1)}_{b}}\right)^{w\alpha^{(1)}_2}\right]^{-1/w}\label{fit2}
\end{eqnarray}
and
\begin{eqnarray}
F_X(t) &=& F_{X,b}^{(2)}\left[\left({t\over
T^{(2)}_{b1}}\right)^{w\alpha^{(2)}_1} + \left({t\over
T^{(2)}_{b1}}\right)^{w\alpha^{(2)}_2}\right.\nonumber\\
&&\left.+\left({T_{b2}^{(2)}\over
T^{(2)}_{b1}}\right)^{w\alpha^{(2)}_2}\left({t\over
T^{(2)}_{b2}}\right)^{w\alpha^{(2)}_3}\right]^{-1/w}\label{fit3},
\end{eqnarray}
to fit the selected one-break and two-break LCs, respectively, we
can obtain the slopes ($\alpha$), the break times ($T_b$), and the
X-ray fluxes at the first break ($F_{X,b}$) of the LCs, as listed in
Tables 1 and 2. The distributions of these fitting parameters are
exhibited in Figure 4 (solid histogram), which shows that (i)
$\alpha_1^{(1)}$ is statistically a bit larger than
$\alpha_1^{(2)}$, and (ii) $\alpha_2^{(1)}$ could be usually larger
than $\sim 1.2-1.4$, whereas $\alpha_2^{(2)}$ is inclined to be
smaller than $\sim 1.2-1.4$, as expected by our model. But when we
use Equation (\ref{fit2}) to fit only the first two segments of the
two-break LCs, the difference between $\alpha_2^{(1)}$ and
$\alpha_2^{(2)}$ becomes ambiguous as shown by the dashed open
histogram in Figure 4. Anyway, in contrast to the centralization of
$T_{b2}^{(2)}$ at $\sim10^{4-5}$ s, the obvious lack at least before
$\sim10^{6}$ s of the second break in the one-break LCs still
suggests a clear difference between these two types of LCs, even
though a longer-term ($>10^6$ s) observation may be able to detect a
second break in some so-called one-break afterglows.

For the above observational results, a plausible explanation can in
principle be provided by our model where the one- and two-break
afterglows are respectively attributed to the shock and wind
emission, although it seems hard to observationally judge (within
the present small sample) whether there is an intrinsic difference
between the one- and two-break LCs. On one hand, for the temporal
decay indices, the values of $\alpha_{1,2}^{(1)}$ and
$\alpha_{1,2,3}^{(2)}$ are statistically in rough agreement with the
model-predicted $\alpha_{1,2}^{(\rm sh)}$ and $\alpha_{1,2,3}^{(\rm
mw)}$, respectively. On the other hand, the similar distributions of
$T_{b}^{(1)}$ and $T_{b1}^{(2)}$ can also be naturally understood by
ascribing both of them to a same physical origin, i.e., the {\it
r}-mode instability requiring $T_{b}^{(1)}\sim T_{b1}^{(2)}\sim
(1+z)T_g$. Of course, for a more comprehensive observational test,
an investigation on the afterglow spectral information is desired.
On the theoretical aspect, it would be useful to derive some closure
relations between the temporal decay index and the spectral index in
the wind emission model, as did by some authors for the shock model
(Zhang et al. 2006; Liang et al. 2006; Willingale et al. 2007).
Unfortunately, such an attempt now is restricted by the ignorance of
the specific radiation mechanisms of the winds.

\subsection{Implications to the central magnetars}
Following the above comparison between the model and the
observations, here we simply connect the selected one- and two-break
afterglows with the shock- and wind-dominated emission,
respectively, in order to give a sight into the properties of the
possible existing central magnetars. To be specific, for the
two-break afterglows, the values of the characteristic timescales in
the model can be easily determined by $T_g\sim T_{b1}^{(2)}/(1+z)$,
$T_c\sim T_{b2}^{(2)}/(1+z)$, and $T_m = (T_gT_c)^{1/2}$ (for $q =
1$). Then the spin periods and magnetic field strengths of the
magnetars can be easily obtained by
\begin{eqnarray}
P_{i,-3}&=&a^{-1/6}T_{g,3}^{1/6},\label{p-3}\\
B_{14}&=&1.4P_{i,-3}T_{m,5}^{-1/2}.\label{b141}
\end{eqnarray}
However, for the one-break afterglows, the values of $T_c$ and thus
$T_m$ can not be found from the observations. So we have to seek
help from Equations (\ref{lmd}), (\ref{lxsh0}), and $\mathcal
L_{X,\rm ei}^{\rm sh}\sim \mathcal L_{X,b}^{\rm obs}$.\footnote{The
isotropically-equivalent observational X-ray luminosity at
$T_b^{(1)}$ can be calculated by $\mathcal L^{\rm obs}_{X,b} = 4\pi
d_l^2 F_{X,b}^{(1)}$, where the luminosity distance reads $d_l(z) =
{(1+z)c \over H_0} \int^z_0 [\Omega_\Lambda+\Omega_m(1 +
z')^3]^{-1/2} dz'$ with cosmological parameters $\Omega_\Lambda=
0.73$, $\Omega_m= 0.27$, and $H_0 = 73~\rm km~ s^{-1}Mpc^{-1}$.}
Then we get
\begin{eqnarray}
B_{14}\approx \left[\left(\mathcal L_{X,b,46}^{\rm
obs}\right)^{4/(3p-2)}P_{i,-3}^2\right]^{1/2},\label{b142}
\end{eqnarray}
with model parameters
$\nu_{X,17.5}=g_{p,-0.5}=\epsilon_{e,-1}=\epsilon_{B,-2}= \mathcal
E_{i,50}=f_{B,-1}=1$ and $\zeta\sim1$. Using the above equations, we
derive the values of the magnetar parameters $P_i$ and $B$ from both
the one-break and two-break samples, as listed in Tables 1 and 2,
respectively. Figure 5 further shows that, for the model parameters
adopted, there is no clear separation between the two samples.
However, keep in mind that the magnetar parameters inferred from the
one-break afterglows are strongly sensitive to the uncertain model
parameters (especially $p$, $\epsilon_e$ and $f_{B}$).

So as a conservative treatment, here we only analyze the relatively
credible magnetar parameters from the two-break afterglows. First,
the hypothesis of $B < B_c$ is favored by the inferred magnetic
field strengths as $B\sim10^{14-15}$ G. Secondly, by calculating the
magnetic dipole luminosity using Equation (\ref{lmd}), we can
estimate the X-ray radiation efficiency of the magnetar winds, which
reads
\begin{eqnarray}
\xi=f_B\mathcal L_{X,b}^{\rm obs}/L_{\rm md,i}\sim(0.01-0.1)f_{B,-1}
\end{eqnarray}
as shown in Table 2. Finally, we even find that the magnetic field
strengths and the spin periods of the magnetars may satisfy a loose
correlation as $B \propto P_i^{-3/2}$. This is qualitatively in
agreement with the dynamo origin of the magnetic fields (e.g., Xu et
al. 2001).

\section{Summary and discussion}
Based on two assumptions of that (i) some GRB central objects are
millisecond magnetars and (ii) the magnetar winds can continuously
produce X-ray emission whose luminosity tracks the magnetic dipole
luminosity, we investigate the temporal behaviors of the GRB X-ray
afterglows arising from an emitting magnetar wind and an energized
external shock together. The competition between the internal- and
external-origin emission components determines the diversity of the
observed X-ray afterglow LCs. A comparison between the model and
observations shows that the model-predicted shock- and
wind-dominated emission is qualitatively consistent with the
observed one-break and two-break afterglows, respectively.

In the conventional shock model, the second break of the two-break
afterglows is always connected to the jet break, which is however
seriously challenged by the usually observed chromatic breaks (Liang
et al. 2008).
In contrast, the chromatic breaks could be acceptable for the
internal-origin emission. On one hand, such an argument is supported
by the lack of the optical counterparts of X-ray flares, which are
of internal origin. On the other hand, in view of the possible small
radii where the internal dissipations occur, the low-frequency
emission of the wind is quite likely to be suppressed, for example,
by some self absorption effects. Therefore, for some GRBs, while
their X-ray afterglows are contributed by the magnetar winds, the
optical emission could be still dominated by the external shocks. In
this case, chromatic breaks would be detected naturally.

In this paper, we mainly concern the ordinary GRB X-ray afterglows
that may be associated with a magnetar with a relatively low
magnetic field ($B<B_c$). For simplicity we do not pay much
attention to some unusual X-ray afterglows such as the afterglows
behaving as a plateau followed by a steep decay (e.g., GRBs 060607A
and 070110). As discussed in Section 3, such X-ray afterglows could
be internal afterglows produced by the magnetars with relatively
high magnetic fields ($B>B_c$). In this case, however, if a low
value of $B$ is found, the related magnetar could be a candidate of
strange quark stars rather than neutron stars, since only strange
stars can suppress the {\it r}-mode instability effectively (Yu et
al. 2009b).

\acknowledgements This work made use of data supplied by the UK
Swift Science Data Centre at the University of Leicester. We
acknowledge useful comments by the referee. This work is supported
by the GRF Grants of the Government of the Hong Kong SAR under
HKU7011/09P. YWY is also supported by the Self-Determined Research
Funds of CCNU (CCNU09A01020) from the colleges' basic research and
operation of MOE.

\begin{table}
\caption{The fitting parameters of the one-break X-ray afterglow LCs
and the corresponding magnetar parameters.}
\begin{tabular}{ c  c c c c c  c c c}
\hline \hline
 GRB    & $z^{\dag}$ & $\alpha^{(1)}_1$ & $\alpha^{(1)}_2$  & $T^{(1)}_{b}$ & $F_{X,b}^{(1)}$   &  $P_i^\ddag$  &    $B^\ddag$
  \\     
 &&&& ($10^3$ s)& ($10^{-11}\rm ~erg~ s^{-1}cm^{-1}$) &  ($10^{-3}$ s)  &  ($10^{14}$ G)\\
 \hline
 051109B &   0.08  &   0.14 &   1.20 &   2.34 &   0.41 &  1.14  &  0.20    \\
  060604 &   2.68  &   0.09 &   1.26 &  14.68 &   0.34 &  1.26  &  4.68    \\
  060714 &   2.711 &   0.01 &   1.24 &   2.23 &   3.01 &  0.92  &  5.55    \\
  060729 &   0.54  &   0.28 &   1.42 &  88.33 &   0.93 &  1.96  &  3.87    \\
  060906 &   3.686 &   0.29 &   1.80 &  12.83 &   0.24 &  1.18  &  4.81    \\
 060923A &$<$2.8   &   0.10 &   1.25 &   4.00 &   1.10 &  1.00  &  4.78    \\
 080905B &   2.374 &   0.20 &   1.46 &   3.49 &   8.52 &  1.00  &  8.64    \\
  090418 &   1.608 &   0.30 &   1.61 &   2.43 &   9.91 &  0.99  &  6.21    \\
  090423 &   8.26  &  -0.16 &   1.43 &   4.35 &   1.02 &  0.88  &  9.00    \\
  091018 &   0.971 &   0.29 &   1.23 &   0.50 &  24.35 &  0.80  &  3.53    \\
\hline
\end{tabular}
\\
$^{\dag}$The GRB redshifts are taken from the website
http://www.mpe.mpg.de/$\sim$jcg/grbgen.html\\
$^\ddag$The values here are obtained with model parameters
$\nu_{X,17.5}=g_{p,-0.5}=\epsilon_{e,-1}=\epsilon_{B,-2}= \mathcal
E_{i,50}=f_{B,-1}=1$ and $\zeta\sim1$.
\end{table}

\begin{table}
\caption{The fitting parameters$^*$ of the two-break X-ray afterglow
LCs and the corresponding magnetar parameters.}
\begin{tabular}{ c  c c c c c  c c  c c c c c}
\hline \hline
 GRB    & $z^{\dag}$ & $\alpha^{(2)}_1$ & $\alpha^{(2)}_2$ & $\alpha^{(2)}_3$ & $T^{(2)}_{b1}$ &  $T^{(2)}_{b2}$&$F_{X,b}^{(2)}$    &  $P_i^\ddag$  &    $B^\ddag$    &      $\xi /f_{B,-1}$       \\
 &&&&&($10^3$ s)& ($10^5$ s)& ($10^{-11}\rm ~erg~ s^{-1}cm^{-1}$)             &  ($10^{-3}$ s)  &  ($10^{14}$ G)  &                \\
 \hline
 050315  &   1.949   &  0.00 (0.00)  & 0.74 (0.77) &  2.11   &  9.63 (10.35) &  2.82   &  0.99 (0.98)  &     1.22  &   4.06  &    0.035         \\
 050505  &   4.27    & -0.05 (0.01)  & 1.14 (1.26) &  1.92   &  5.60 (6.64)  &  0.52   &  2.77 (2.48)  &     1.01  &   7.86  &    0.082         \\
 060614  &   0.125   &  0.01 (0.02)  & 1.40 (1.59) &  2.32   & 37.00 (38.90) &  1.42   &  0.70 (0.68)  &     1.79  &   3.12  &    0.0003        \\
 060807  & $<$3.4    & -0.10 (-0.10) & 1.24 (1.25) &  2.40   &  4.74 (4.74)  &  0.59   &  2.14 (2.13)  &     1.01  &   7.27  &    0.044         \\
 070521  &   1.35    &  0.11 (0.12)  & 1.20 (1.32) &  2.40   &  1.03 (1.26)  &  0.23   & 16.86 (15.94) &     0.87  &   8.48  &    0.014         \\
 080310  &   2.42    &  0.05 (0.12)  & 1.19 (1.25) &  2.22   &  5.41 (6.34)  &  0.92   &  1.28 (1.15)  &     1.08  &   5.92  &    0.022         \\
\hline
\end{tabular}\\
$^*$The numbers in the  brackets are obtained by fitting the first
two segments by Equati on (\ref{fit2}).
\\
$^{\dag}$The GRB redshifts are taken from the website
http://www.mpe.mpg.de/$\sim$jcg/grbgen.html\\
$^\ddag$The values are inferred with $a\sim1$.
\end{table}

\begin{figure}
\resizebox{\hsize}{!}{\includegraphics{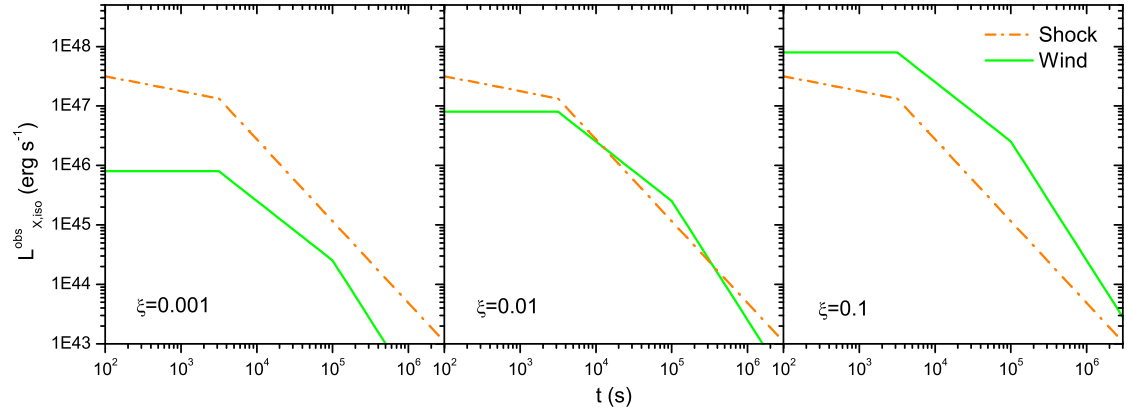}}
\caption{Illustrative X-ray ($\nu_X=3\times 10^{17}$ Hz) afterglow
LCs contributed by the magnetar wind with varying $\xi$ as labeled
and by the energized shock with $\epsilon_e=0.2$ and
$g_{p,-0.5}=\epsilon_{b,-2}= E_{i,50}=1$. The magnetar parameters
are taken to be $P_{i}=1.2$ ms, $B=4\times10^{14}$ G, and $f_B=0.1$.
}
\end{figure}
\begin{figure}
\resizebox{\hsize}{!}{\includegraphics{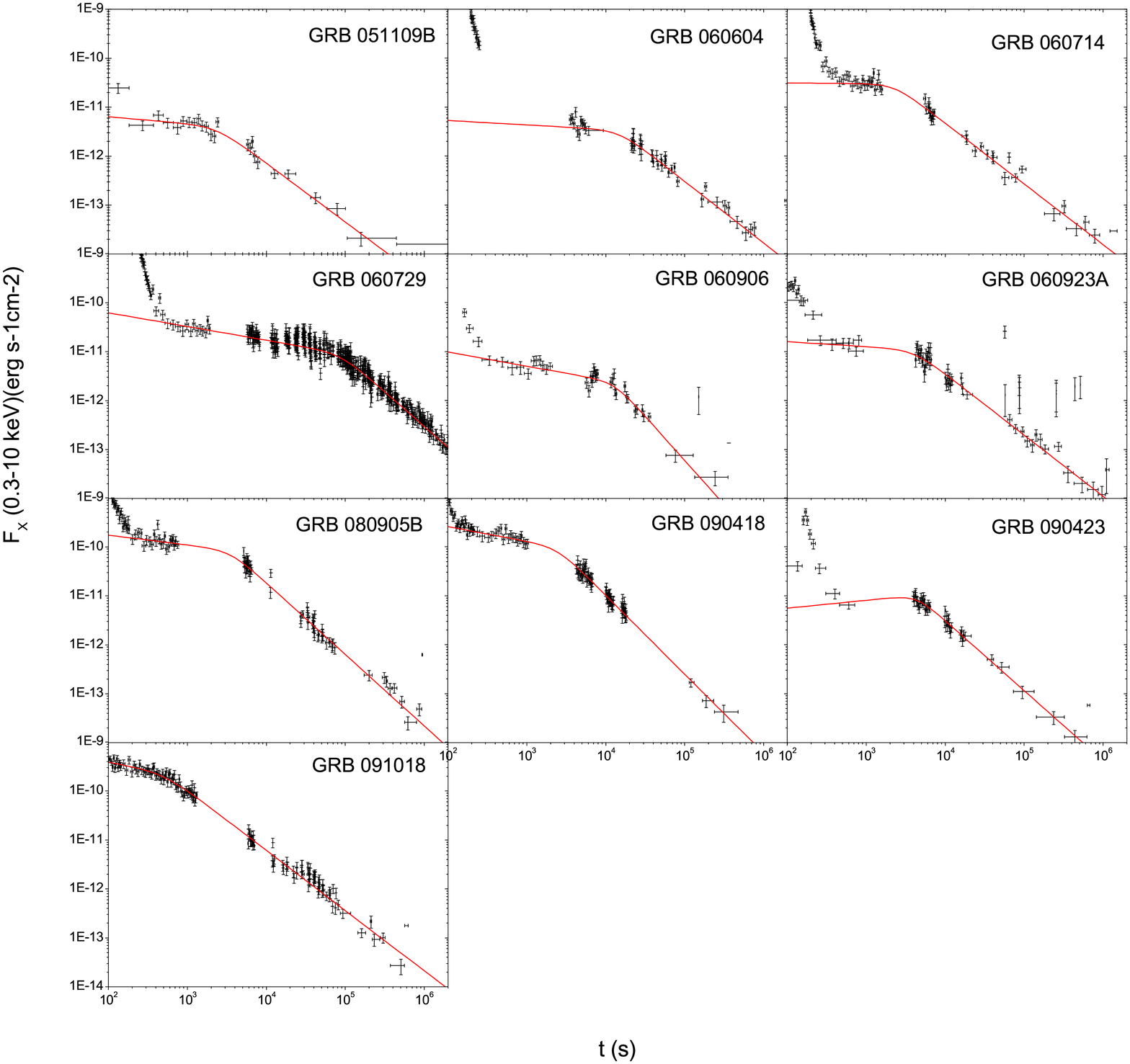}} \caption{Selected
one-break X-ray afterglow LCs, which can be fitted by Equation
(\ref{fit2}) as shown by the solid lines.}
\end{figure}
\begin{figure}
\resizebox{\hsize}{!}{\includegraphics{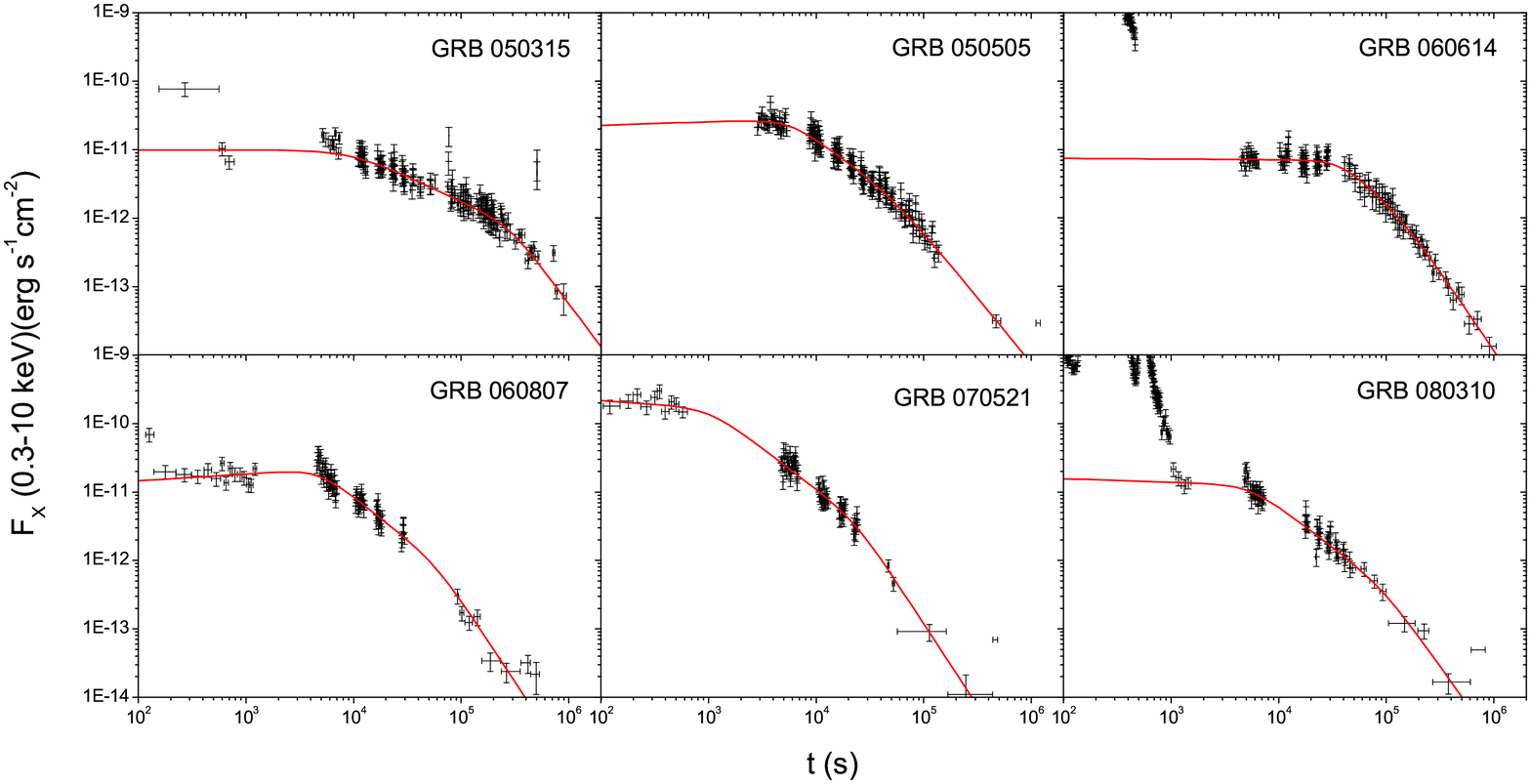}} \caption{Selected
two-break X-ray afterglow LCs, which can be fitted by Equation
(\ref{fit3}) as shown by the solid lines.
}
\end{figure}
\begin{figure}
\resizebox{\hsize}{!}{\includegraphics{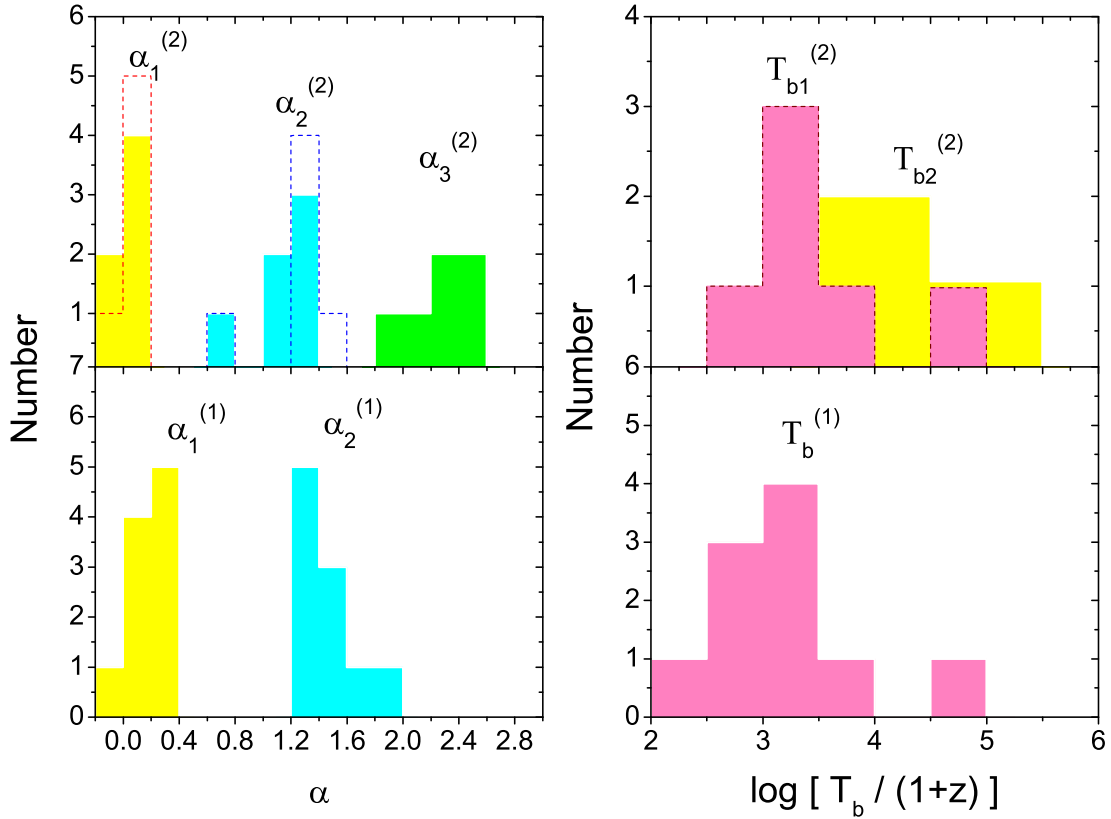}}
\caption{Distributions of the fitting parameters of the one-break
(lower panel) and two-break (upper panel) X-ray afterglow LCs (solid
histogram),  which are respectively fitted by Equations (\ref{fit2})
and (\ref{fit3}). The dashed open histogram shows the results of the
fitting to the first two segments of the LCs by Equation
(\ref{fit2}).}
\end{figure}
\begin{figure}
\resizebox{0.5\hsize}{!}{\includegraphics{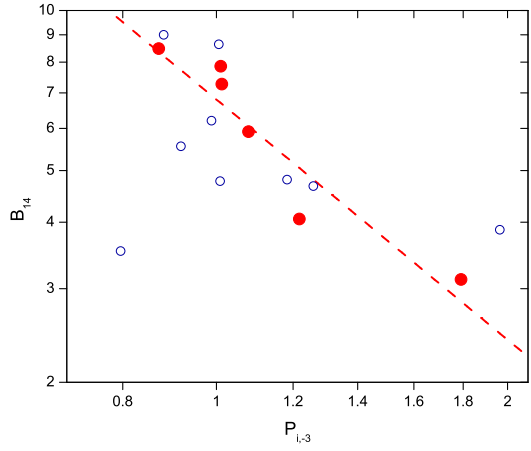}}
\caption{Distribution in the $P_i-B$ panel of the magnetars inferred
from both the one-break (open cycle) and two-break (solid cycle)
X-ray afterglows. The dashed line represents
$B_{14}\sim7P_{i,-3}^{-3/2}$.}
\end{figure}

\end{document}